\documentclass[multphys,vecphys]{svmult}

\usepackage{makeidx}         % allows index generation
\usepackage{graphicx}        % standard LaTeX graphics tool
\usepackage{multicol}        % used for the two-column index
\usepackage[bottom]{footmisc}% places footnotes at page bottom
\makeindex             % used for the subject index
\begin{document}
\def\3801{NGC\,3801\index{NGC 3801}}
\def\cena{Cen A\index{Centaurus A}}
\def\1052{NGC\,1052\index{NGC 1052}}

\title*{Shock heating by nearby AGN}
\author{J.H. Croston\inst{1}\and
R.P. Kraft\inst{2}\and
M.J. Hardcastle\inst{1}
}
\institute{School of Physics, Astronomy and Mathematics, University of
Hertfordshire, College Lane, Hatfield AL10 9AB, UK
\texttt{j.h.croston@herts.ac.uk}
\and Harvard-Smithsonian Center for Astrophysics, 60 Garden Street,
Cambridge, MA~02138, USA}
\maketitle

\section{Abstract}

Shock heating by radio jets is potentially an important process in a
range of environments as it will increase the entropy of the heated
gas. Although this process is expected to occur in the most
powerful radio-loud AGN, strong shocks have so far only been detected
in nearby low-power radio galaxies. Here we discuss X-ray detections
of strong shocks in nearby galaxies, including a new detection of
shocked gas around both lobes of the nearby radio galaxy
NGC\,3801\index{NGC 3801} with inferred Mach numbers of 3 -- 6 and a total
injected energy comparable to the thermal energy of the ISM within 11
kpc. We discuss possible links between shock heating, AGN fuelling and
galaxy mergers and the role of this type of system in feedback models.

\section{When and where is shock heating important?}
\label{intro}

%Although AGN feedback is now understood to play an important role in
%the formation and evolution of galaxies and in determining the
%observational properties of galaxy clusters, the details of which
%energy transfer mechanisms operate, and in what circumstances, remain
%uncertain. Despite the widespread expectation that supersonic
%expansion of powerful radio galaxies should produce shock heating,
%direct observational evidence for this process has been difficult
%
%
%Shock heating by radio jets has been the focus of many
%analytical models and hydrodynamical simulations (e.g. Voit \& Donahue
%2005; Zanni et al. 2005). However, most of
%the observational evidence to date for heating of the intracluster
%medium and radio jet/environment interactions has shown that gentler
%heating processes and mixing may be more important at least at the
%centres of clusters. Despite the widespread expectation that
%supersonic expansion of powerful radio galaxies should produce shock
%heating, direct evidence for this process has been difficult to find.
%It is therefore important to investigate more carefully what
%environmental conditions might be expected to produce shock heating.
%
All radio-loud AGN, whatever their radio luminosity and eventual
morphology, are expected to go through an initial phase of supersonic
expansion before coming into pressure balance (see, for example,
\cite{hei98}). The length of this phase, the amount of energy injected
into the external medium during this phase, and the location of the
energy injection depend on the jet power and density of the
environment. In a poor environment, the radio source will remain
overpressured for longer, so that the shock heating phase will be
longer lived. It therefore seems likely that the two places where
shock heating will be easiest to detect are in the poorest
environments, and in the environments of the most powerful AGN.
Indeed, the first (and until recently only) direct detection of
radio-galaxy shock heating was in the galaxy halo of the nearest radio
galaxy Centaurus A \cite{kra03}, which has a low radio luminosity and
FRI morphology, but whose inner lobes are still in the supersonic
expansion phase (see Section~\ref{cenas}).

More recently weak shocks have been detected in the cluster
environments of several more powerful radio galaxies, e.g. M87
\cite{for05}, and the FRII sources Cygnus A \cite{wil06,bel06} and
Hydra A \cite{nul05}; however, there remains no convincing case of a
strong shock associated with an FRII radio galaxy. In addition, FRII
radio galaxies for which measurements exist of both the internal
pressure (via lobe inverse Compton emission) and the external pressure
appear to be close to pressure balance rather than strongly
overpressured \cite{h02,c04,bel04}, so that lobe expansion is not
likely to be highly supersonic. While it is not possible to rule out
an important role for strong shocks produced by powerful radio
galaxies, the observational evidence suggests that it is in the early
stages of radio-source evolution, for both FRI and FRII sources, that
shock heating is most important. Although the main emphasis of
most work on radio-source impact has been on the group and cluster
scale effects of radio galaxies, the impact of shock heating on the
ISM of AGN host galaxies is likely to be dramatic, as we demonstrate
below.

In the following sections we review the first detection of
radio-source shock heating in Centaurus A before presenting a new
example of strongly shocked gas shells in the ISM of \3801 that
share some characteristics with \cena\, but also show some important
differences. Finally, based on the nuclear properties and host galaxy
characteristics of systems with detected strong shock heating, we
discuss the links between shock heating and AGN fuelling and possible
implications for the role of shock heating in feedback models.

\section{\cena\ and \1052}
\label{cenas}
\begin{figure}
\centering
\includegraphics[width=12.0cm]{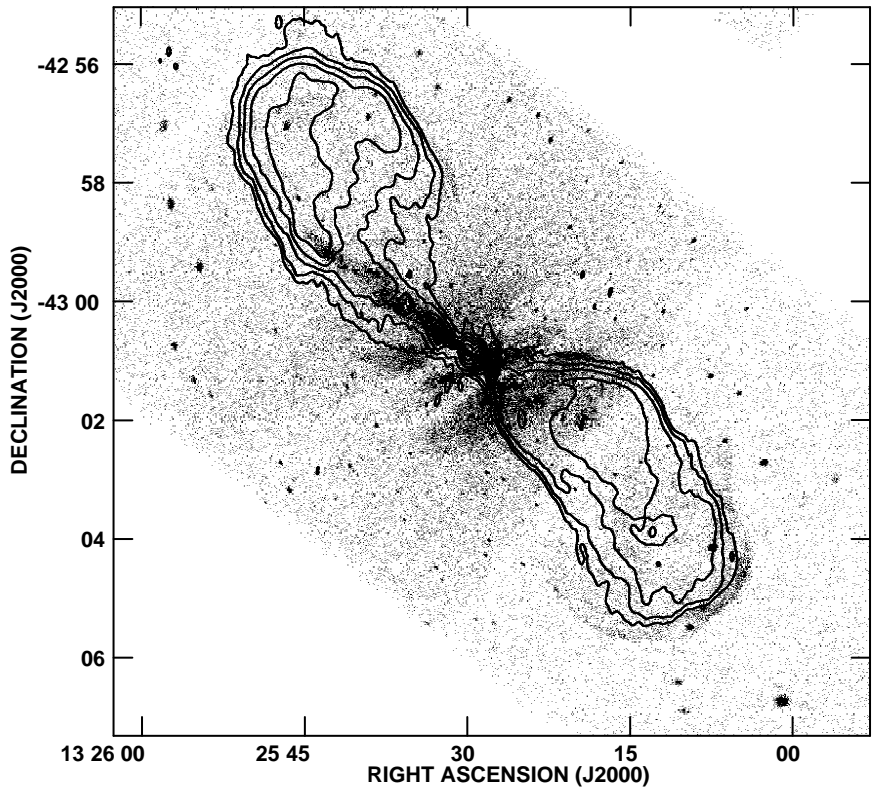}
\caption{The shell of shocked gas around the south-western lobe of
  Centaurus A as observed with {\it Chandra} \cite{kra03,kra06}. 20cm
  radio contours \cite{h06} are overlaid.}
\label{cena}
\end{figure}
Kraft et al. (2003) \cite{kra03} detected a bright shell of hot gas
surrounding the south-west inner lobe of the nearest radio galaxy
Centaurus A. Fig.~\ref{cena} shows more recent {\it Chandra} data
\cite{kra06} illustrating the sharp X-ray shell. The shell has a
temperature ten times higher than that of the surrounding interstellar
medium, and the total thermal energy of the shell is a significant
fraction of the energy of the ISM. Centaurus A has an FRI morphology,
so would traditionally have been expected to have subsonically
expanding lobes; the detection of strongly shocked gas in this system has
highlighted the fact that energy input via shocks is likely to be
important in the early stages of expansion for all types of radio
galaxies.
\begin{figure}
\centering
\vbox{
\includegraphics[width=12.0cm]{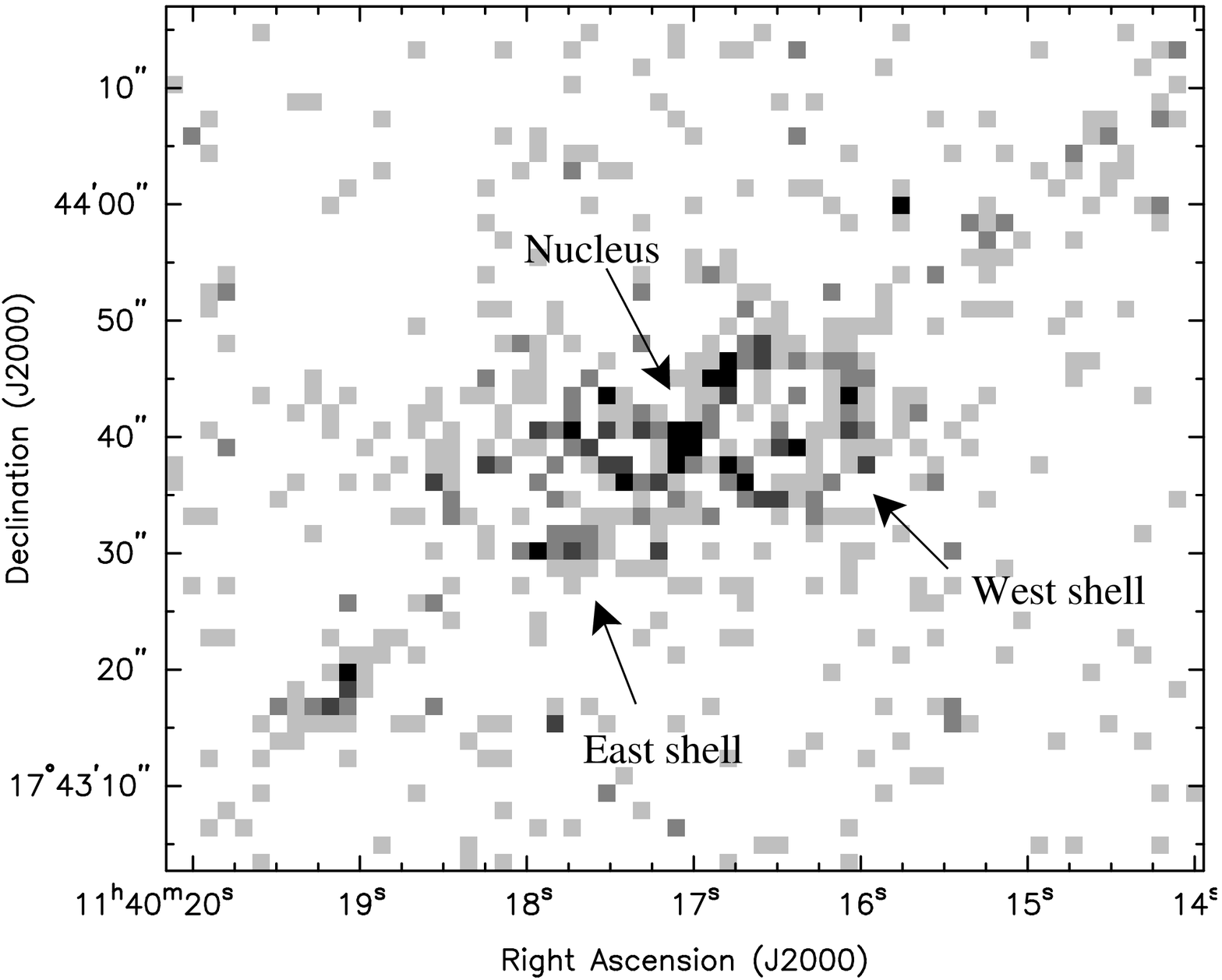}
\includegraphics[width=12.0cm]{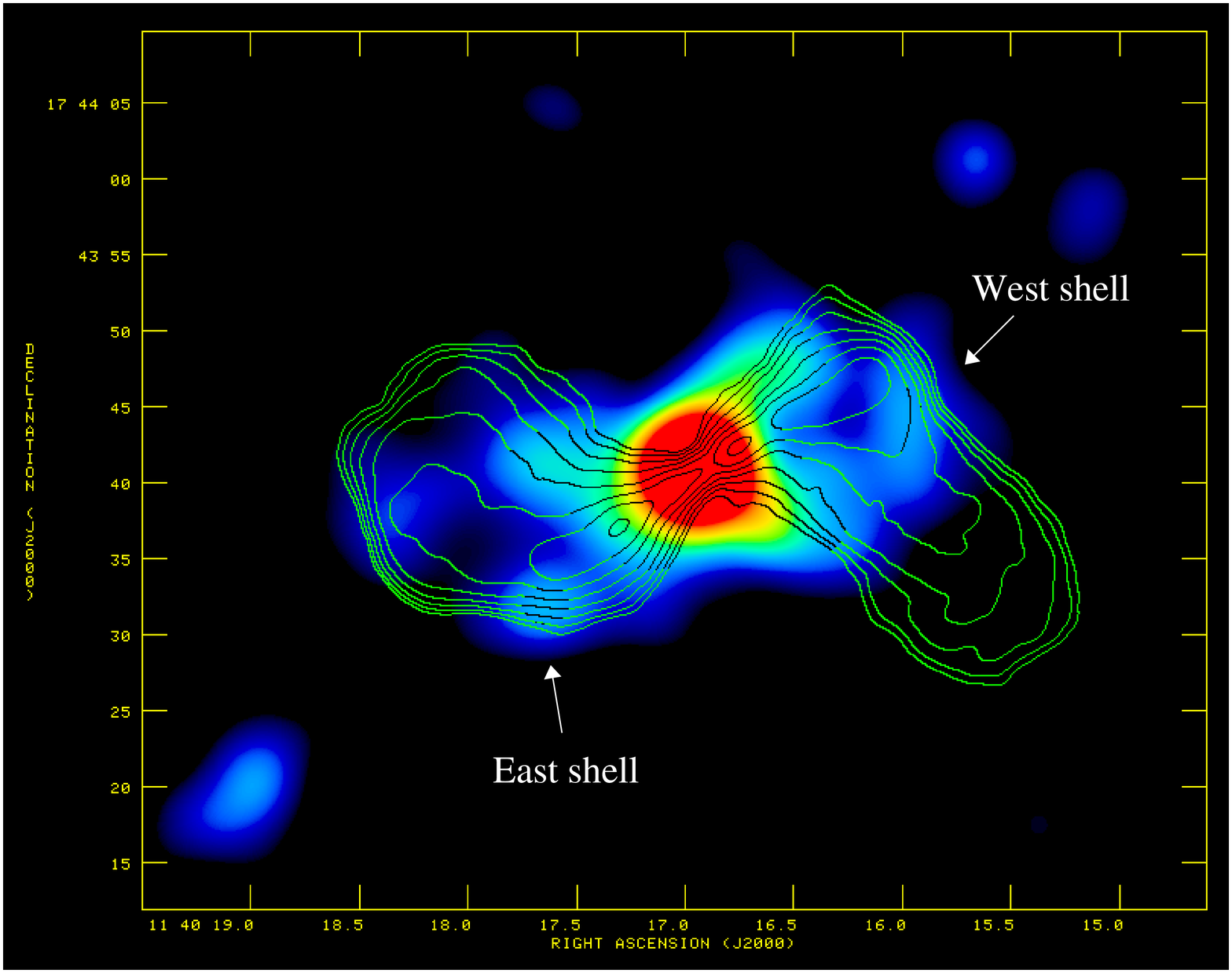}}
\caption{The shocked gas shells of \3801 as observed by {\it
    Chandra}. Top: binned 0.5 - 2 keV counts; bottom: Gaussian
    smoothed 0.5 - 2 keV image with 20 cm radio contours overlaid from new
    VLA data.}
\label{3801}
\end{figure}

\1052 is another nearby galaxy where it has been suggested that
the small radio source could be shocking and heating its hot ISM
\cite{kad04,c05}. A recent, deep {\it Chandra} observation reveals in
detail the radio-related X-ray structure hinted at by the earlier
snapshot observation, but does not show clear evidence for shocked
shells, suggesting that the radio-source/environment interaction in
this system may be considerably more complex than the shock heating
seen in \cena.

\section{A new detection of shock heating in \3801}
\label{3801s}
Our recent {\it Chandra} observations of \3801 \cite{c06} revealed
a second definite example of strong shocks produced by a small FRI
source on galaxy scales. Fig~\ref{3801} shows the {\it
Chandra}-detected emission from \3801, which traces well the outer
edges of the radio lobes.

We can rule out a non-thermal model for the X-ray emission based on
its spectrum, and find best-fitting {\it mekal} temperatures of 1.0
keV and 0.7 keV for the West and East lobes, respectively. The
undisturbed interstellar medium has a temperature of 0.23 keV. We find
that the observed density contrast is consistent with the value of 4
expected for a strong shock, using the mean properties of the shell
and the ISM density halfway along the lobe. The shells are
overpressured by a factor of 13 - 20 and the shell pressure is $\sim
7$ times the synchrotron minimum internal lobe pressure (consistent
with the general finding that FRI minimum pressures are typically an
order of magnitude lower than external pressure
\cite{mor88,wb00,c03}).

We estimated the shock Mach number using two methods, as descibed in
more detail in \cite{c06}: applying the Rankine-Hugoniot jump
conditions using the observed temperature jump gives ${\cal M} \sim 3
- 4$; alternatively, ram pressure balance gives ${\cal M} \sim 5 - 6$.
The discrepancy between the two methods is probably due to the
expected temperature and density structure of the shell and the
interstellar medium (in both cases the data are too poor to constrain
these). Nevertheless this is a clear detection of strongly shocked gas
with ${\cal M} \sim 3 - 6$, which implies a lobe expansion speed of
$\sim 600 - 1200$ km s$^{-1}$.

The total thermal energy stored in the hot gas shells is $\sim 8
\times 10^{55}$ ergs, and for ${\cal M} \sim 4$, the total kinetic
energy of the shells is $\sim 9 \times 10^{55}$ ergs. The total energy
of the shells, $1.7 \times 10^{56}$ ergs is comparable to $P_{int}V$,
the approximate total energy available from the radio source as work;
however, it is $\sim 25$ times the minimum work required to inflate
the lobe cavities ($\sim 7 \times 10^{54}$ ergs), so that a simple
calculation of the radio-source energy input from the cavity size
would be a significant underestimate. The total energy is also
equivalent to the thermal energy of the ISM within 11 kpc (or 25
percent of the thermal energy within 30 kpc). Shock heating is
therefore the dominant energy transfer mechanism during this phase of
radio-source activity, and will have dramatic long term effects: part
or all of the ISM may be expelled from the galaxy, and the entropy of
the gas will be permanently increased. The internal energy of the
radio source ($\sim 4 \times 10^{56}$ ergs) must also eventually be
transferred to the environment.

The age of the radio source in \3801 is estimated to be $\sim 2
\times 10^{6}$ y from radio spectral ageing and dynamical arguments,
which implies an energy injection rate of $\sim 3 \times 10^{42}$ ergs
s$^{-1}$. This should correspond to a considerable fraction of the jet
power, which is consistent with a rough estimate of its jet power
based on scaling that of 3C\,31 \cite{lb02} by the ratio of radio
luminosities of \3801 and 3C\,31. The rate of mechanical energy
extracted is roughly an order of magnitude higher than the
accretion-related X-ray luminosity, so that the AGN is more
efficiently converting energy into jet production than radiation. We
also find that the Bondi accretion rate from hot gas would be
sufficient to power this radio outburst, for $\eta \sim 0.05$.

\section{A link between shock heating and AGN fuelling?}

Both \cena\ and \3801 are disturbed ellipticals with evidence for
fairly recent mergers. Another property that the two sources have in
common is that their nuclear X-ray spectra show a component of
emission with heavy intrinsic absorption (N$_{H} > 5 \times 10^{22}$
cm$^{-2}$ in both cases) as seen in high-excitation FRII radio-galaxy
X-ray spectra \cite{hec06}. This is in constrast to the vast majority
of FRI radio galaxies, which possess no direct evidence for
accretion-related X-ray emission or a torus \cite{eva05,hec06}. It is
therefore interesting to speculate that these systems represent a
particular class of FRI radio outburst fuelled by cold gas that may be
driven into the centre during gas-rich mergers, a mechanism that is
unlikely to operate in rich group or cluster-centre FRI radio sources.
If this is true, then the shock heating process is not
self-regulating, as most of the AGN energy goes into the hot phase of
the ISM, so that the accretion rate of cold material is not directly
affected. \cena\ and \3801 may represent a class of systems at the
massive end of the galaxy luminosity function that experience extreme
heating effects.

\section{Conclusions}

We have recently found a second example of strong shocks associated
with the radio lobes of a nearby galaxy \cite{c06}, with a total
energy in the shock-heated shells $\sim 25$ times the minimum that
would have been required to inflate the cavities subsonically: shock
heating is therefore the dominant energy transfer mechanism for this
source. Young radio galaxies should all go through an early
stage of supersonic expansion, and the examples of \cena\ and \3801
show that this stage can have dramatic effects on the host galaxy ISM.
As this stage is comparatively short-lived, and outbursts of the
luminosity of \3801 and \cena\ are currently only detectable to $z
\sim 0.04$ in the radio, further examples of this process may be
difficult to find with current generation instruments; however, they
are expected to be orders of magnitude more common than Cygnus A type
radio outbursts. The nuclear and host galaxy properties of \3801
and \cena\ suggest that the shock heating in these galaxies may be
directly related to their merger history; we suggest that
merger-triggered radio outbursts could be an important galaxy feedback
mechanism.

%
%\begin{table}
%\begin{tabular}{lll}
%\hline\noalign{\smallskip}
%\end{tabular}
%\end{table}

%\includegraphics[height=4cm]{figure.eps}
%\picplace{5cm}{2cm} % Give the correct figure height and width in cm
%\end{figure}
%
% For built-in environments use
%
%
%
% BibTeX users please use
% \bibliographystyle{}
% \bibliography{}
%
% Non-BibTeX users please follow the syntax
% the syntax of "referenc.tex" for your own citations

%%%%%%%%%%%%%%%%%%%%%%%%%%%%%%%%%%%%%%%%%%%%%%%%%%%%%%%%%%%%%%%%%%%%%%  }

%%%%%%%%%%%%%%%%%%%%%%%%%%%%%%%%%%%%%%%%%%%%%%%%%%%%%%%%%%%%%%%%%%%%%%

\printindex
\end{document}